\documentclass[aps,pra,twocolumn,groupedaddress,showpacs]{revtex4}
\usepackage[latin9]{inputenc}
\usepackage{amssymb}
\usepackage{graphicx}
\usepackage{amsmath}
\usepackage{color}
\usepackage{mathrsfs}
\usepackage{float}
\usepackage{indentfirst}
\usepackage{mathrsfs}
\usepackage{float}
\usepackage{indentfirst}
\usepackage{textcomp}

\newcommand\PLOTFILE[1]  {./#1}

\newcommand\bk        {\mathbf{k}} 
\newcommand\br        {\mathbf{r}} 
\newcommand{\vx}{{\mathbf x}}

\newcommand{\COMMENTED}[1]{}

\begin{document}
\title{Exact Ground-State Properties of Strongly Interacting Fermi Gases \\ in Two Dimensions}
\author{Hao Shi}
\author{Simone Chiesa}
\author{Shiwei Zhang}

\affiliation{Department of Physics,
             The College of William and Mary,
             Williamsburg, Virginia 23187}

\begin{abstract}
Exact calculations are performed on the two-dimensional
strongly interacting, unpolarized, uniform Fermi gas with a zero-range attractive interaction. 
Two  auxiliary-field approaches are employed which accelerate the sampling of imaginary-time paths 
using BCS trial wave functions and a force bias technique. Their combination   
enables calculations 
on large enough lattices to reliably compute ground-state properties in the thermodynamic limit. 
A new equation of state is obtained, with a parametrization provided, which can serve as a benchmark and 
allow accurate  comparisons with experiments. 
The pressure, contact parameter, and condensate fraction are determined 
     systematically vs.~$k_F a$. The momentum distribution, pairing correlation, and the structure of the
     pair wave function are computed. 
     The use of force bias to accelerate the Metropolis sampling of auxiliary-fields in determinantal approaches is 
     discussed.

\end{abstract}

\pacs{03.75.Ss, 05.30.Fk, 02.70.Ss, 03.75.Hh}

\maketitle

Exact results on fundamental models are uncommon, especially for strongly interacting fermion 
systems. In the rare cases where they exist
(for example in one-dimensional models by Bethe ansatz or density matrix renormalization group \cite{DMRGPRL1992,DMRGRMP2005}),
they have invariably played an integral role in bringing about physical insights, advancing our understanding, 
and serving as benchmarks for the development of new theoretical and computational approaches. 

The Fermi gas with a zero-range attractive interactions is a model for strongly interacting fermions which 
has generated a great deal of research activities \cite{RevModPhys.80.1215, RevModPhys.80.885}. 
The model is of interest in both condensed matter
and nuclear physics. As a model it is rather unique in that, 
thanks to advances in experimental techniques using ultracold atoms, it  
can be realized in a laboratory with great precision and control \cite{fermi-gas-2008,RevModPhys.80.885}.

In three-dimensions (3D) the interplay between 
experiment, theory and computation has lead to rapid advances \cite{LeLuo2009,Ku03022012,PhysRevA.84.061602,PhysRevA.79.013627}. 
An example is seen in the evolution \cite{PhysRevA.87.023615} of the determination of the so-called Bertsch parameter at unitarity.
Quantitative comparisons have allowed validation of our understanding and provided an impetus for 
developments of both experimental and theoretical techniques.
The remarkable level of agreement achieved recently
between calculation \cite{PhysRevA.84.061602} and experiment \cite{Ku03022012} 
demonstrates the tremendous progress
towards precise understanding and control of strongly correlated quantum matter.

The two-dimensional (2D) Fermi gas has attracted considerable recent interest \cite{PhysRevA.67.031601, PhysRevLett.96.040404,
PhysRevA.78.043617, PhysRevLett.95.170407, PhysRevA.77.063613, PhysRevB.75.184526, PhysRevA.77.053617, PhysRevLett.106.110403,Huihu-2D-2011}, especially with 
its experimental realization using highly anisotropic trapping potentials \cite{PhysRevLett.105.030404}. 
In 2D a bound state always exists, and the BCS-BEC cross-over offers rich possibilities between
the interplay of  inter-particle spacing (density) and interaction strength, where effects beyond the mean-field description will be
more pronounced than in 3D.
Interest in this model is further enhanced by the 2D nature of many of the most 
interesting and complex materials, including high-$T_c$ cuprate superconductors and topological superconductors \cite{PhysRevLett.103.020401}.

In this paper, we obtain \emph{exact} numerical results on  
the ground state of the strongly interacting 2D spin-balanced uniform 
Fermi gas.
To date the most accurate numerical results on the 2D system have mainly come from diffusion Monte Carlo (DMC) simulations
\cite{PhysRevLett.106.110403}.
These calculations, however, involve the fixed-node approximation \cite{anderson1976quantum,DMCCeperley} and lead to
systematic errors 
which are difficult to estimate; furthermore, some of the correlation functions 
that are central to the physics of these systems
are not readily available from DMC.
Here, we employ  
two auxiliary-field 
quantum Monte Carlo (AFQMC) approaches: one based on the branching random walk method used 
in the 3D study in Ref.~\cite{PhysRevA.84.061602},
and the other a novel approach 
in the Metropolis path-integral framework which dramatically improves efficiency.
Their combination allows us to calculate the thermodynamics and pairing properties  
exactly in the entire range of interaction strengths.
 
Our calculations are performed on periodic
lattices. 
We use supercells of 
up to 3,000 sites,
containing about $120$ particles, with projection length in imaginary time of $\beta>50$ (in unites of $1/E_F$).
For each lattice and Hamiltonian parameters, the calculation is numerically exact, with only statistical 
uncertainties which are fully controlled. 
Systematic extrapolations are then carried out to reach the thermodynamic limit (TL). 

As the interaction in cold atoms is short-ranged compared to the inter-particle spacing,
the uniform 2D Fermi gas can be modeled by a 
lattice 
Hamiltonian 
\begin{equation}
 \hat{H}=t\,\sum_{{\mathbf k},\sigma} \varepsilon_{\mathbf k} c^{\dagger}_{{\mathbf k}\sigma}c_{{\mathbf k}\sigma}+U\sum_{i}^{{\mathcal N}_s} n_{i\uparrow}n_{i\downarrow}\,
 \label{eq:H}
\end{equation} 
with ${\mathcal N}_s=L^2$ sites and
$t=\hbar^2/(2m\Delta^2)$, where $\Delta$ is the lattice parameter. 
Only
the low energy behavior of $\varepsilon_{\mathbf k}$ will be relevant, and
we have used both the Hubbard dispersion $\varepsilon^H_{\mathbf k}=4-2(\cos k_x+\cos k_y)$ and the quadratic dispersion 
$\varepsilon^q_{\mathbf k}=k^2_x+k^2_y$. 
In this form, the momentum $k_x$ (or $k_y$) is defined on the lattice, with 
units $2\pi/L$, and $k_x\in [-\pi,\pi)$.
The on-site interaction is attractive and is given by \cite{PhysRevA.86.013626}
\begin{equation}
\frac{U}{t}=-\frac{4\pi}{\ln(k_Fa)-\ln({\mathcal C} \sqrt{n})}\,,
 \label{eq:map-U2kFa}
\end{equation}
which is tuned, 
for each lattice density $n \equiv N/{\mathcal N}_s$ and Fermi momentum $k_F=\sqrt{2\pi n}/\Delta$,
to produce the desired 
2D scattering length $a$, defined as the position of the node of the zero-energy $s$-wave solution of the two-body problem. The constant ${\mathcal C}$ in Eq.~(\ref{eq:map-U2kFa}) 
depends on the dispersion relation: 
${\mathcal C}^H = 0.49758$ and ${\mathcal C}^q= 0.80261$.

We employ two AFQMC methods to study this model: the branching random walk approach, 
and an accelerated Metropolis approach
 with a force bias. In the first \cite{PhysRevA.84.061602}, we
 project the ground-state wave function by importance-sampled random walks in Slater determinant space \cite{shiweiPRL1995,PhysRevLett.90.136401}.
 A BCS wave function,
taken from the solution of the gap equation for the same discretized Hamiltonian,
is chosen as the trial wave function,
and the mixed estimator \cite{PhysRevA.84.061602,AFQMC-lecture-notes-2013} is used to calculate the ground-state energy.
The BCS trial wave function 
shortens the convergence time in the imaginary-time projection, 
and greatly reduces the
Monte Carlo statistical fluctuations, as illustrated in the 3D case \cite{PhysRevA.84.061602}.

Our second approach is based on the 
ground-state path integral form of AFQMC, but introduces several 
advances, including accelerated sampling (described in more detail in Appendix~\ref{appendix:fb}) by 
a dynamic force bias \cite{AFQMC-lecture-notes-2013}, which enables global moves 
of fields on a time slice with acceptance ratio of over 90\%, and
control of the Monte Carlo variance \cite{HShi-SZhang-to-be-published}. 
Its main advantage over the the open-ended branching random walk
approach is the ease with which any observables can be computed, 
and we use it to compute the momentum distribution and
correlation functions.
(Since there is no sign problem here, no constraint is needed, which is the primary motivation for 
using the open-ended branching random walk form.) 
With this approach, our calculations typically have $\beta\sim 320$ or larger (in units of $t^{-1}$),
discretized with 
over 12,800 time-slices.

These technical advances result in orders of magnitude improvement in sampling efficiency,
which makes it possible to achieve the high numerical 
accuracy 
presented in this work.
In both approaches, the computational cost scales as  $\sim {\mathcal N}_s N^2 \beta$.
The linear scaling with 
$ {\mathcal N}_s$ is important, as it enables calculations on large lattice sizes.
To approach the TL, 
we first extrapolate calculations 
to the continuum limit by taking
${\mathcal N}_s \rightarrow\infty$ while holding  $N$  fixed.
The number of particles, $N$, is then increased   
until convergence is reached within our statistical 
accuracy, as illustrated next. 

\begin{figure}[htbp]
  \includegraphics[scale=0.6]{\PLOTFILE{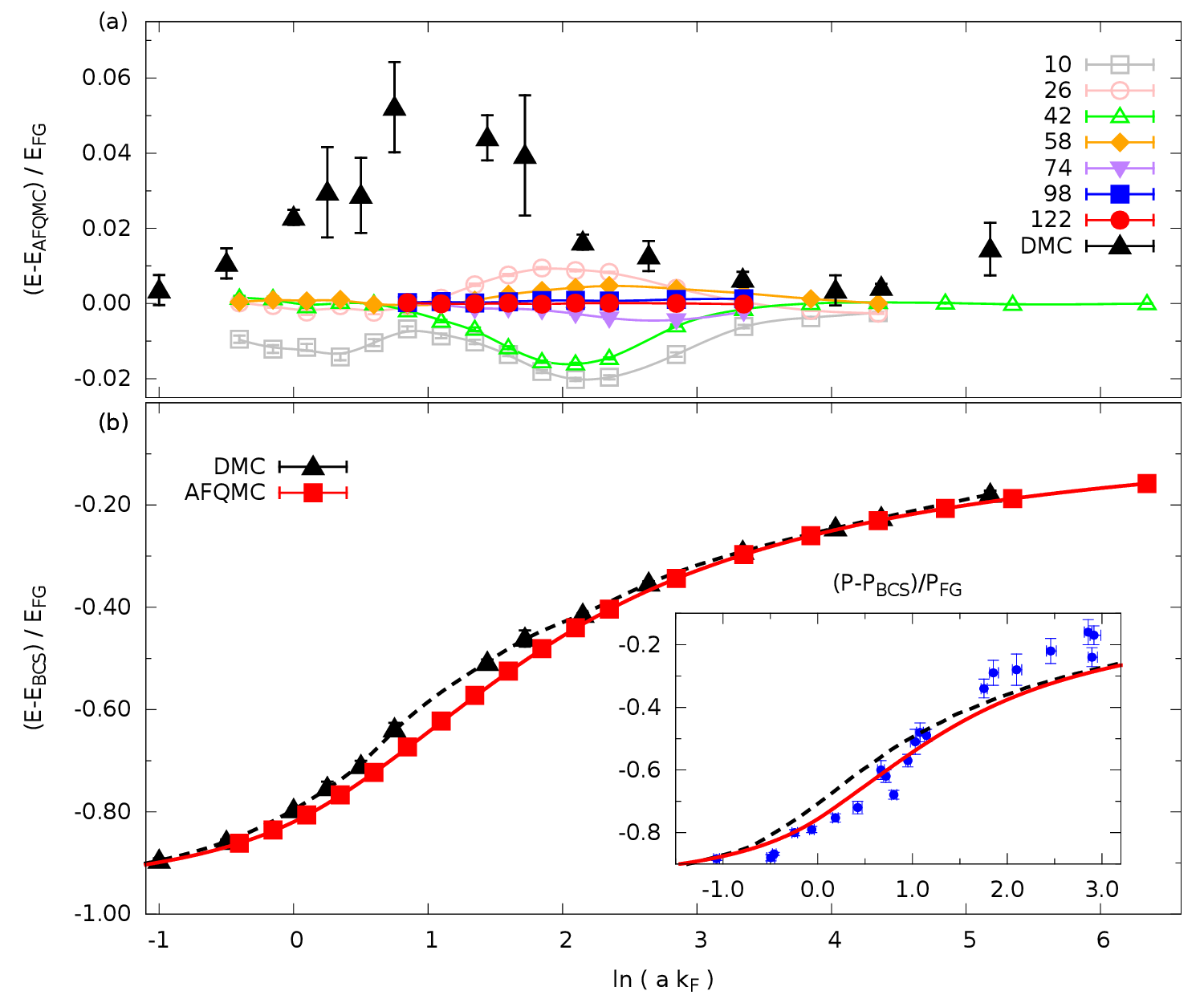}}
  \caption{\label{fig:energy} (Color online)
  Calculated equation of state. 
  The top panel shows the energy, relative to the final AFQMC results, for finite number of particles, $N$.
  Also shown are the DMC results of Ref.~\cite{PhysRevLett.106.110403}, which are variational. 
  Note the small scale of the vertical axis.     
  The bottom panel shows the AFQMC (and DMC) results at the TL, relative to the BCS result.
  A fit has been performed on the AFQMC results for the EOS. The result is given 
  in Eqs.~(\ref{eqn:fitbcs}-\ref{eqn:fitbec}) and shown as the solid line. The inset in panel (b) compares 
  the calculated pressure from AFQMC (solid line) and DMC (dashed, taken from Ref.~\cite{PhysRevLett.112.045301})  with 
  experiment \cite{PhysRevLett.112.045301} (points) in the crossover region.
}
\end{figure}

Figure~\ref{fig:energy} displays the calculated equation of state (EOS), in units of the Fermi
gas energy $E_{\rm FG}=\pi n t$,
as a function of
the interacting strength, 
$x\equiv \ln (k_F a)$.
A table of the AFQMC data can be found in Appendix~ \ref{appendix:eos}.
The top panel illustrates the convergence to the TL,
where AFQMC energies are shown 
for fixed $N$. At each $x$, 
the energy has been extrapolated to the continuum limit, 
using a 
4th-order polynomial in $1/L$. 
In the more strongly interacting cases, we take advantage of the fact that 
$\varepsilon^q_{\mathbf k}$ and $\varepsilon^H_{\mathbf k}$ 
produce energies which 
converge to a common limit from opposite directions and perform both sets of calculations to reduce the uncertainty in 
the extrapolation. In the opposite regime, energies from the quadratic dispersion shows less dependence on $L$ 
and they are used alone. 
We illustrate the extrapolation procedure in Appendix~\ref{appendix:fss}.
The error bar of each symbol, barely noticeable in the graph, combines
the QMC statistical error (negligible) at each
$L$ and a conservative estimate of the uncertainty from the extrapolation,
which typically involves half a dozen or more data points from each dispersion relation, with 
$L$
ranging from $\sim 15$ to $45$  (and larger if necessary).

The results for different values of $N$ show that convergence is reached  to within 
our statistical accuracy
by $N\sim 100$ \cite{FS-corr-note}. This is consistent with 
DMC results \cite{PhysRevLett.106.110403} 
        which observed no significant change between $N$ of $26$ and $98$. 
	The DMC results provide the current best estimate of the EOS and are included in Fig.~1. We see that the 
	error from the fixed-node approximation is
	largest in the crossover region, at intermediate values of $x$. 
	The maximum error is about 10\% of the ``correlation energy'', 
	the difference between the BCS and exact energies. 
	
	In addition to serving as a benchmark for theory, 
	the new EOS can provide validation for experiments.
	Experiments are fast developing;  
	in 3D remarkable precision \cite{Ku03022012} 
	was reached in the measurement of the Bertsch parameter
	(with uncertainties only slightly larger than  our symbol size in the top panel of Fig.~1).
       In the inset in the bottom panel, we show a comparison of the calculated pressure with the latest experiment in 
       2D  \cite{PhysRevLett.112.045301}. In the crossover regime, better agreement with experiment is seen  
       with the new result than 
       with DMC. There may be other factors contributing to the discrepancy between experiment and theory
       \cite{PhysRevLett.112.135302,2014arXiv1408.2737L}. We leave more detailed comparisons of our 
       results and experiment to a future publication.
        
\begin{table*}
\caption{\label{tab:fitparam} 
Final parameter values (full digits in Appendix \ref{appendix:fitparam}) in the parametrization [Eqs.~(\ref{eqn:fitcross}-\ref{eqn:fitbec})] of the exact EOS from QMC. 
}
\begin{ruledtabular}
\begin{tabular}{ccccccccc}
$i$ & 0 & 1& 2 & 3 & 4 & 5 & 6 & 7\\
\hline
$a^l_i$&-11.8041 & 14.6755 & -4.85508\\
$a_i$& -0.81984 & 0.12733 & 0.06851& -0.01451 & -0.00919 & 0.00419 & -0.00064 & 3.4312$\times 10^{-5}$\\
$a^r_i$& & & -0.06085 & 0.36401 & -0.61531
\end{tabular}
\end{ruledtabular}
\end{table*}

	We parametrize the computed EOS by 
	$E_c\equiv E_\text{QMC} -E_\text{BCS}$
[note that $E_{\rm BCS}/E_{\rm FG}$ is related to the two-body binding energy by
$1-\epsilon_B/(2E_{\rm FG})$, and is given by $1-8e^{-2(\gamma+x)}$
where $\gamma=0.57721$ is Euler's constant]:

\begin{displaymath}
 \frac{E_c}{E_{\rm FG}}= \left \{
\begin{array}{ll}
 f^l (x),& x\leq 0.2664\,;\\
 f(x),  &  0.2664<x<4.3058\,;\\
 f^r (x),& x\geq 4.3058\,. 
\end{array}
\right.
\end{displaymath}
The intermediate region is fitted with a 7th-order polynomial
\begin{equation}
\label{eqn:fitcross}
f (x)=\sum_{i=0}^7 a_{i} x^{i}\,.
\end{equation}
In the BCS region, the form is based on perturbative results \cite{PhysRevB.45.10135,PhysRevA.90.053633} 
\begin{equation}
\label{eqn:fitbcs}
f^r (x)=-\frac{1}{x}+\sum_{i=2} ^4\frac{a^r_i}{x^i},,
\end{equation}
while 
in the BEC regime a dimer form is used
\begin{equation}
\label{eqn:fitbec}
f^l (x) = -1+\frac{0.5}{X} \bigg[1-\frac{\ln(X)}{X}+\frac{c_1}{X} 
 +\frac{\sum_{i=0}^2 a^l_i (\ln X)^i}{X^2}\bigg]\,, 
\end{equation}
where $X\equiv c_0-2x$ with $c_0=3.703$
from the dimer scattering length $\sim0.557a$ given by few-body calculations  \cite{PhysRevA.67.031601},
and $c_1=\ln(\pi)+2\gamma+0.5$.
The parameters in Eqs.~(\ref{eqn:fitbcs}) and (\ref{eqn:fitbec})
 are determined by 
continuity conditions (value and first two derivatives) from Eq.~(\ref{eqn:fitcross}). The parameters and
the locations of the transition between different regions are then varied in a small range to further 
minimize the variance of the overall fit with the QMC data. 
The final 
parameters are listed in Table~\ref{tab:fitparam} \cite{Param-EOS-BCS-a2--note}.

\begin{figure}[htbp]
  \includegraphics[scale=0.6]{\PLOTFILE{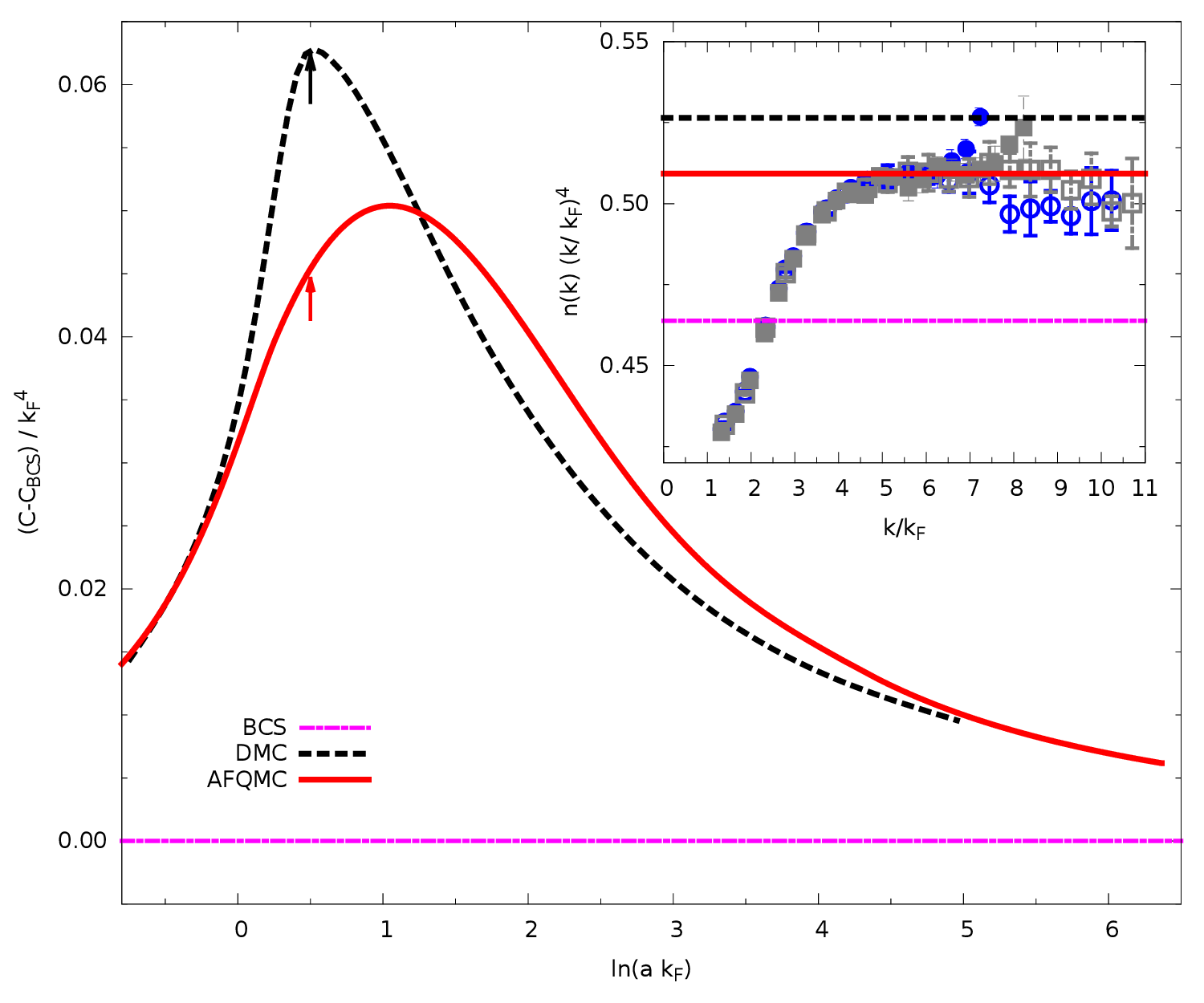}}
  \caption{\label{fig:contact} (Color online) 
  The contact parameter $C$. The main figure shows the result of $C$ 
  (relative to the BCS result)
obtained from  Eq.~(\ref{eqn:C-EOS}). The statistical uncertainty is smaller than the line thickness. 
  DMC \cite{PhysRevLett.106.110403} and BCS results are also shown for comparison.
 The inset shows $n(\bk) k^4$ vs $k\equiv |\bk|$  at $x=0.5$.
The horizontal 
lines give 
the $C$ values from DMC, AFQMC and BCS (top to bottom), indicated by the arrows in the main figure.
The $n(\bk)$ data are from two systems, with $L=45$ (circles) and $51$ (squares), 
respectively, and $N=58$. Results are plotted for $\bk$ along both the horizontal 
(solid symbols)
and diagonal (open) directions.
  }
\end{figure}

\begin{figure*}[htbp]
  \includegraphics[scale=0.6]{\PLOTFILE{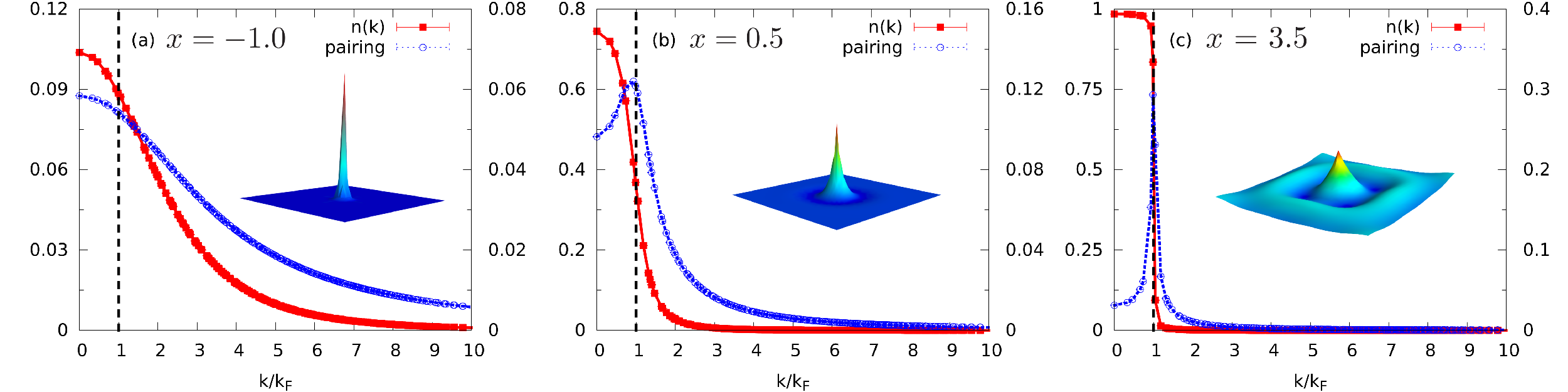}}
  \caption{
   (Color online) 
  Momentum distribution and pair 
  wave functions in three regimes of interaction strengths, $x\equiv \ln(a k_F)$.
  In each panel, the vertical tick labels on the left are for $n(\bk)$ and those on the right are for 
  $\phi_{\uparrow\downarrow}(\bk)$, both plotted vs.~$k$ (in units of $k_F$). 
  Note the different scales between the three panels. 
  The inset shows the real-space 
  wave function
  $\psi_{\uparrow\downarrow}({\mathbf r})$ vs.~${\mathbf r}$ in a 3D plot. The lattice has
  ${\mathcal N}_s=2025$ sites, with density $n=0.0286$. 
  }
  \label{fig:pairwf}
\end{figure*}

The contact \cite{Tan20082952,Tan20082971} is important to the physics of dilute gases,
and can potentially be measured experimentally \cite{PhysRevLett.109.220402,PhysRevLett.108.145305}.
With the functional form of the EOS, it is straightforward to determine the contact:
\begin{equation}
\label{eqn:C-EOS}
\frac{C}{k^4_F}=\frac{1}{4} \frac{d(E/E_{\rm FG})}{dx}\,.
\end{equation}
The result is shown in Fig.~\ref{fig:contact}.
An alternative approach to obtain the contact parameter is from the 
tail of the momentum distribution \cite{Tan20082971,PhysRevLett.106.205302}: $n(\bk) k^4 \rightarrow C$ at large $k$. This provides an internal check on the consistency and accuracy of the calculation.
As illustrated in the inset, a clear plateau is present 
before edge effects start to manifest as $k$ approaches the cut-off value, giving a $C$ value in 
excellent agreement with that from the EOS.
(The full momentum distribution $n(\bk)$ is shown in Fig.~\ref{fig:pairwf} for three representive
interaction strengths.)
The pressure and the chemical potential can be obtained from 
simple combinations of the energy and contact: 
$P/P_{\rm FG}=2\,C/k_F^4 + E/E_{\rm FG}$, which was applied in the inset in Fig.~\ref{fig:energy},
and $\mu/\mu_{\rm FG}=C/k_F^4 + E/E_{\rm FG}$. 

We next quantify how the pairing properties evolve as a function of interaction strength.
The zero-momentum pairing matrix (of dimension ${\mathcal N}_s\times {\mathcal N}_s$),
\begin{equation}
 M_{\bk\bk^\prime}=\langle \varDelta_\bk^{\dagger}\varDelta_{\bk^\prime}\rangle-
 \delta_{\bk\bk'}\langle c^{\dagger}_{\bk\uparrow} c_{\bk\uparrow}\rangle
 \langle c^{\dagger}_{-\bk\downarrow} c_{-\bk\downarrow} \rangle\,,
\end{equation}
is computed in the many-body ground state,
where the 
pair creation operator 
$\varDelta_\bk^{\dagger}\equiv c^{\dagger}_{\bk\uparrow} c^{\dagger}_{-\bk\downarrow}$.
We associate \cite{RevModPhys.34.694} the leading eigenstate 
with the pair wave function in $\bk$-space,
$\phi_{\uparrow\downarrow}(\bk)$.
This is shown in Fig.~\ref{fig:pairwf} 
for three characteristic interaction strengths. 
The inset shows the corresponding real-space structures, 
$\psi_{\uparrow\downarrow}({\mathbf r})$,
obtained from the 
 Fourier transform of $\phi_{\uparrow\downarrow}(\bk)$. 
 In the BEC regime, the momentum distribution is very broad, the pair 
 wave function involves many 
 $\bk$-values, and the pairs are tightly bound like a molecule, as seen in (a). 
 In the BCS regime in (c), on the other hand, modifications to the non-interacting $n(\bk)$ are limited to 
 near the Fermi surface,  with 
 a small number of $\bk$-vectors in its vicinity participating in pairing. The
 pair 
 wave function is sharply peaked near the Fermi surface, and becomes very extended in real space. (Residual finite-size effect can be seen 
 in this case in the second
 ring of  $\psi_{\uparrow\downarrow}({\mathbf r})$ which is affected by the shape of the supercell.)
As $k_Fa$ is increased, the systems crosses over from (a) to (c)  via the  
strongly interacting 
regime represented in (b). 
Beyond the central peak, the wave function $\psi_{\uparrow\downarrow}({\mathbf r})$ in (b)
contains significant radial oscillations, with 
multiple circular nodes. 

\begin{figure}[htbp]
  \includegraphics[scale=0.6]{\PLOTFILE{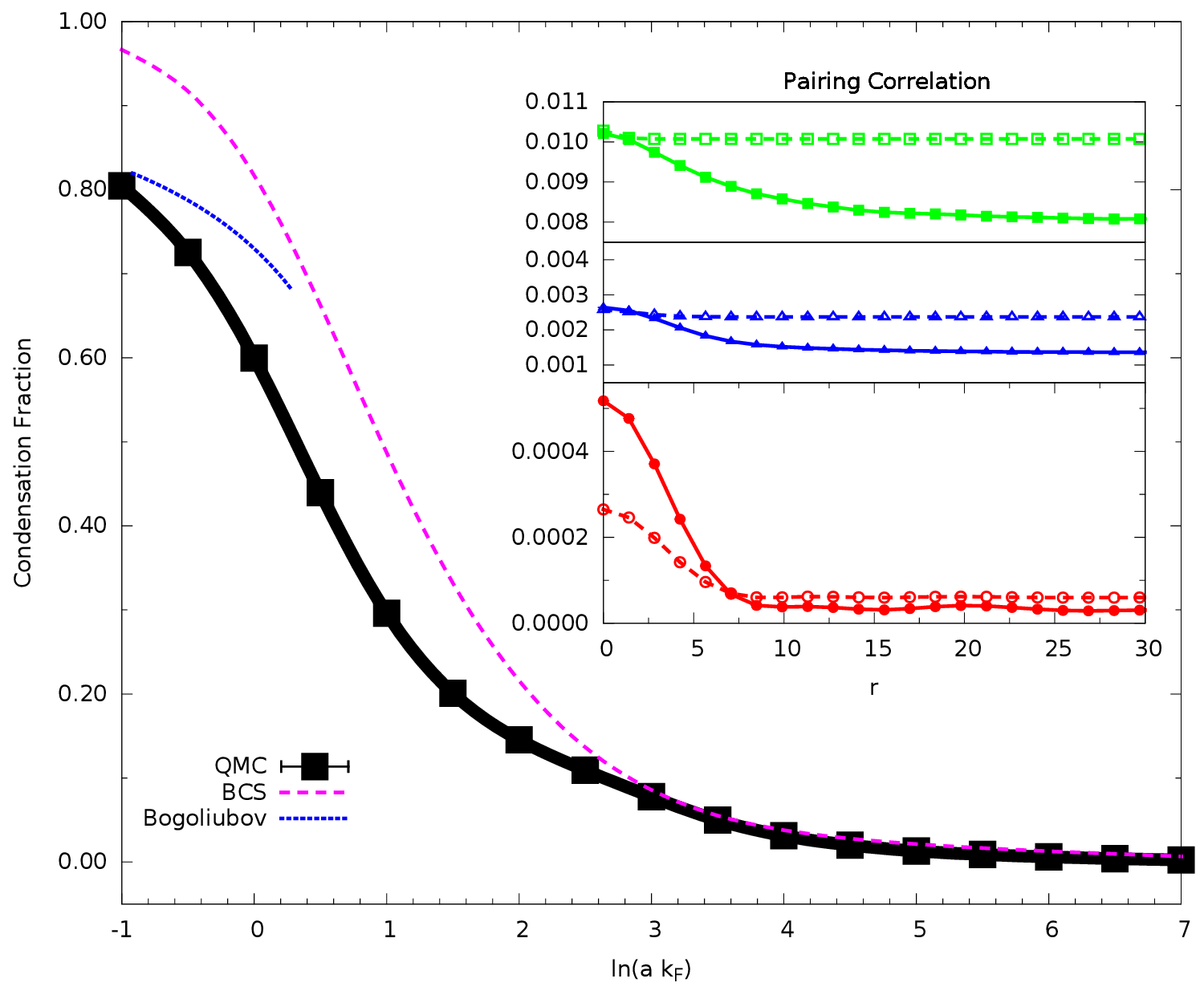}}
  \caption{(Color online)\label{fig:confrac} 
  Condensate fraction and pairing correlation functions.
 In the main graph, the uncertainty in the QMC data (from extrapolation to the TL) is estimated 
 by multiple runs with different sizes and is indicated by the thickness of
   the line. 
   Also shown are BCS results and, in the BEC limit,  Bogoliubov results for Bose gas for reference.
  In the inset,  the pairing correlation function 
  $C(\br)$ is plotted vs.~$r$ for three interaction strengths
  (from top to bottom,  the same parameters as 
  in (a), (b), and (c) of Fig.~\ref{fig:pairwf}).  The dashed lines 
  are from BCS and solid lines are QMC results (error bars smaller than symbol size).
}
\end{figure}

The condensate fraction is given by the  largest
eigenvalue of $M_{\bk\bk^\prime}$
divided by $N/2$. The results are shown in Fig.~\ref{fig:confrac} as a function of interaction.
At the mean-field BCS level $M_{\bk\bk^\prime}=\langle \varDelta^\dagger_\bk\rangle\langle \varDelta_{\bk'}\rangle$, and there is only one non-zero eigenvalue (equal to $\sum_\bk |\langle \varDelta_{\bk'}\rangle|^2)$.
In the many-body ground state,
additional depletion of the condensate is present from scattering into zero-momentum pairs 
distinct from $\phi_{\uparrow\downarrow}(\bk)$.
The BCS condensate fraction and pair wave functions are in reasonable agreement with exact 
results down to $\ln(a k_F) \sim 3$. 
For stronger interactions, the BCS condensate fraction grows significantly faster.
At $\ln (k_F a)\sim -1$, it predicts an essentially $100\%$ condensate as opposed 
to only $80\%$ from the exact result. 
In this regime, Bogoliubov theory of a Bose gas \cite{PhysRevA.3.1067} 
with the dimer scattering length above gives results consistent with the QMC data.
The largest deviation between BCS and exact results occurs in the crossover region, near  
$\ln(a k_F)\sim 0.5$, where the momentum distributions and pair wave functions also
exhibit the largest differences.

We also calculate the real-space on-site 
pairing correlation function:
\begin{equation}
 C({\mathbf r})=\langle c^{\dagger}_{{\mathbf 0} \uparrow} c^{\dagger}_{{\mathbf 0} \downarrow} c_{\br\downarrow}c_{\br\uparrow}\rangle\,,
\end{equation}
where 
the reference point ${\mathbf 0}$ and 
all $\br$ values related by translational symmetry can be averaged over. 
 The results are shown as a function of $r\equiv |\br|$ in the inset in Fig.~\ref{fig:confrac}, 
 for three representative values of interaction strength. 
 Long-range order can be seen in all three regimes, with $C({\mathbf r})$ approaching a finite 
 constant at large $r$.

In summary, we have calculated \emph{exact} properties
of the strongly interacting 2D Fermi gas at zero temperature, 
by a combination of two AFQMC methods.
The equation of state, contact parameter, condensation fraction and pair wave functions
are obtained.  
Improved agreement is seen with the pressure recently measured in quasi-2D experiment compared to
best current (approximate) theoretical results.  
Our results will provide valuable benchmarks for future studies and allow precise comparisons with 
experiments as the latter rapidly develop in 2D. 
The  analytic forms parametrized 
from the accurate numerical results will also facilitate future 
local-density type of calculations \cite{LDA-SF-Bulgac}
in a variety of systems relevant to experiment, including 
thermodynamics and out of equilibrium properties in the presence of 
a trap.
The technical advances in computational techniques,  which
allowed efficient sampling of larger lattices with long imaginary-times and much smaller Monte Carlo variance than
previously possible,
can be expected to have many applications in cold atom systems 
and elsewhere.

\appendix
\section{Generalized Metropolis with force bias \label{appendix:fb}}

In this appendix, we describe our second approach using  the
generalized 
Metropolis procedure to accelerate the sampling of paths in auxiliary field (AF) space. 
We introduce a dynamic force bias, analogous to what is employed in the branching random walk methods
in constrained path or phase-free AFQMC \cite{AFQMC-lecture-notes-2013}, in proposing the updates of the 
field values, which improves the acceptance ratio and hence the MC efficiency.

 To facilitate the description of the sampling algorithm
 we first give a brief sketch of the standard path-integral AFQMC approach, on which more detailed descriptions
can be found in, for example, Refs.~\cite{assaad-lec-note} and \cite{AFQMC-lecture-notes-2013}.
Ground state AFQMC measures the static properties by
\begin{equation}
\label{eqn:projection}
 \langle \hat{O}\rangle =\frac{\langle \psi_T|\,\exp(-\beta \hat{H}/2)\,\hat{O}\,\exp(-\beta\hat{H}/2)\,|\psi_T \rangle}{ \langle \psi_T|\,\exp(-\beta \hat{H})\,|\psi_T \rangle}\,,
\end{equation}
where the Hamiltonian $\hat{H}\equiv \hat{K}+\hat{V}$ is given by Eq.~(\ref{eq:H}). 
We apply the usual Trotter-Suzuki breakup 
\begin{equation}
 e^{-\Delta \tau \hat{H}} \backsimeq e^{-\Delta \tau \hat{K}/2} e^{-\Delta \tau \hat{V}} e^{-\Delta \tau \hat{K}/2}\,
\end{equation}
and the Hubbard-Stratonovich (HS) decomposition 
\cite{PhysRevB.28.4059}
\begin{align}
 e^{\Delta \tau U n_{i\uparrow}n_{i\downarrow}} 
 &=\frac{1}{2} \sum_{x_i=\pm1} e^{(\gamma x_i -\Delta \tau U/2) (n_{i\uparrow} + n_{i\downarrow}-1)} 
 \label{eq:HS-Hirsch-charge} \\
 &\equiv 
\frac{1}{2} \sum_{x_i=\pm1} \hat{b}_i(x_i)\,, \nonumber  
\end{align}
with $ \cosh(\gamma) = \exp( -\Delta \tau U/2 )$,  
arriving at the form 
\begin{equation}
      e^{-\Delta\tau {\hat H}}=
\int d\vx\,  p(\vx) {\hat B}(\vx)\,,
\label{eq:HS}
\end{equation}
where $\vx=\{x_1, x_2, \cdots,x_{{\mathcal N}_s}\}$
The probability density function 
 $p(\vx)$ is uniform for the $2^{{\mathcal N}_s}$ AF configurations under the choice of HS in Eq.~(\ref{eq:HS-Hirsch-charge}), and the one-body propagator
 is ${\hat B}(\vx)= e^{-\Delta \tau \hat{K}/2} \prod_i 
 \hat{b}_i(x_i)\,e^{-\Delta \tau \hat{K}/2} $.
 
  The expression in Eq.~(\ref{eqn:projection}) is then re-written as a path integral of $M\equiv \beta/\Delta\tau$ 
 time slices. Let us consider the $l$-th time slice, and introduce the notation
 \begin{eqnarray*}
\langle\psi_l| & = &
\langle \psi_T|\,
{\hat B}(\vx^{(M)}){\hat B}(\vx^{(M-1)}) \cdots
{\hat B}(\vx^{(l+1)})\,e^{-\Delta \tau \hat{K}/2} \\
|\psi_r\rangle & = &
e^{-\Delta \tau \hat{K}/2}\,{\hat B}(\vx^{(l-1)}){\hat B}(\vx^{(l-2)}) \cdots {\hat B}(\vx^{(1)})
\,|\psi_T\rangle\,,
\end{eqnarray*}
which are both single Slater determinant wave functions if we choose $|\psi_T\rangle$ to be a Slater determinant. 
The integrand of the path-integral in the denominator of Eq.~(\ref{eqn:projection}) then becomes 
\begin{equation}
{\mathcal W}(\vx)=p(\vx)\, \langle \psi_l |  \prod_{i=1}^{{\mathcal N}_s}  \hat{b}_i(x_i) |\psi_r\rangle\,,
\end{equation}
where $\vx$ denotes the collection of AF at time slice $l$.
In the standard way of sampling ${\mathcal W}$, one 
proposes to flip each auxiliary-field $x_i$ one by one, and sweeps through $\vx$. We will update 
the entire configuration $\vx$ (or a sub-cluster of $\vx$ for very large system sizes), simultaneously. We define a force bias  \cite{AFQMC-lecture-notes-2013}:
\begin{equation}
 \bar{n}_{i\sigma}= \frac{ \langle \psi_l |n_{i\sigma} |\psi_r\rangle  } {\langle \psi_l  |\psi_r\rangle}\,,
 \label{eq:FB-dynamic}
\end{equation}
and propose updates of the fields with the probability density: 
\begin{equation}
{\mathcal P}(\vx)\propto p(\vx) \prod_{i=1}^{{\mathcal N}_s} e^{\gamma x_i  (\bar{n}_{i\uparrow} + \bar{n}_{i\downarrow}-1)}\, 
\end{equation}
which can be sampled directly. Detailed balance then leads to a Metropolis acceptance probability given by
\begin{equation}
  {\mathcal A}(\vx\rightarrow \vx') = \min\{1,\frac{{\mathcal W}(\vx^{\prime})\,{\mathcal P}(\vx)}{{\mathcal W}(\vx)\,{\mathcal P}(\vx^{\prime})}\}\,.
\end{equation}
Note that the probability function for proposing transitions
does not depend on the ``current'' configuration of AF, i.e., ${\mathcal P}(\vx\rightarrow \vx')={\mathcal P}(\vx')$.
If ${\mathcal P = W}$, all updates will be accepted. 
Because of the force bias, ${\mathcal P}$ approximates ${\mathcal W}$ up to ${\mathcal O}(\sqrt{\Delta \tau})$, 
leading to typically high acceptance ratio.

Although we have used the discrete charge HS decomposition, 
the algorithm generalizes straightforwardly 
to continuous HS transformations. 
We comment that the use of the dynamic force bias in Eq.~(\ref{eq:FB-dynamic}) effectively introduces a 
background subtraction \cite{PhysRevB.88.125132,AFQMC-lecture-notes-2013} in the decomposition of  Eq.~(\ref{eq:HS-Hirsch-charge}). 
That is, if one were to employ the standard updating algorithms
\emph{without the force bias}, one would find Eq.~(\ref{eq:HS-Hirsch-charge})
much less efficient than a continuous charge decomposition which subtracts a constant 
background. This discrepancy in efficiency grows more as the system density decreases, which is especially
relevant since
the systems studied here are at the low density limit. (See Ref.~\cite{PhysRevB.88.125132}
for an analysis of the efficiency of HS transformations, and Ref.~\cite{PhysRevE.70.056702} for 
discussion on how the dynamic force bias automatically introduces an optimal constant background shift.)

Some other features of our algorithm are:

\begin{itemize} 

\item Since we always work in the dilute limit, the memory is saved by only storing the wave function and calculating the Green function on the fly. We divide the path of $M$ slices into 
$\sqrt{M}$ blocks, and
only track one block each time. The wave function at the beginning of each block is stored. The largest number of wave functions stored in our code is $\sim2\sqrt{M}$.

\item The wave function is transformed
between real and momentum space by fast Fourier transformation, so that all the one-body operators during projection are diagonal, and Green functions in different space are easily obtained.

\item When we only need the energy, we separate it into kinetic and potential energy.  They are diagonal either in momentum or real space,  where we do not need to calculated the whole Green function.
To improve statistics, we measure the energy
anywhere along the path  and combine them, including the mixed estimator on both side.

\item The standard determinantal QMC formalism as sketched above turns out to have a divergence of the Monte Carlo variance. We discuss the variance problem and its solution 
separately elsewhere \cite{HShi-SZhang-to-be-published}. The solution involves the introduction of a bridge link, which we 
have implemented in the calculations presented here.
The force bias and basic sampling algorithm described above
remain unchanged.

\end{itemize}

\section{Extrapolation to the continuum limit \label{appendix:fss}}
We have described the extrapolation procedure of our lattice results to the continuum limit, and the subsequent 
analysis to reach the thermodynamic limit. Here
we illustrate the finite  size extrapolation in few-body systems.

\begin{figure*}[htbp]
  \includegraphics[scale=0.5]{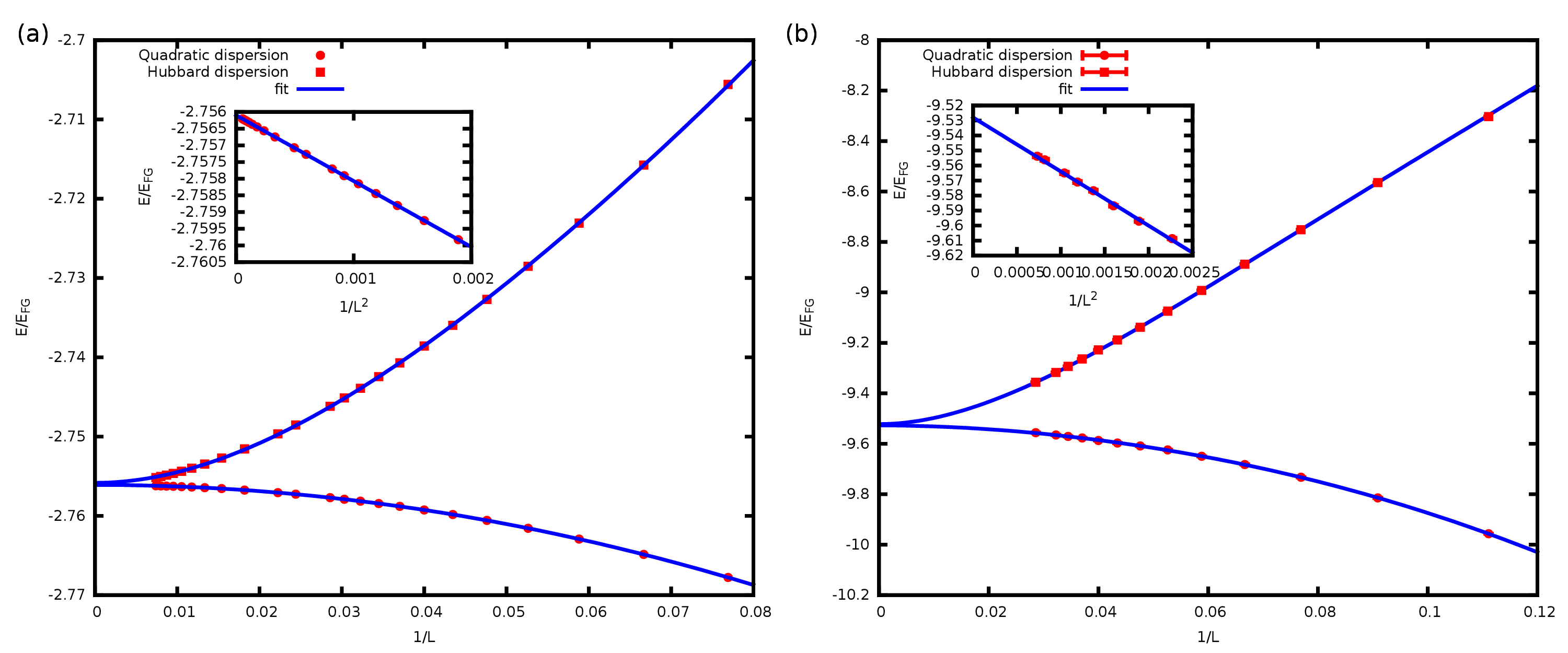}
  \caption{\label{fig:fss} (Color online) 
  Extrapolation of finite-size lattice to the continuum limit in few body problems. Panel (a) shows 
  exact diagonalization results for the two-body problem at $\ln(a k_F)=0.5$, while panel (b) shows
  QMC solutions for the four-body problem at $\ln(a k_F)=0.0$.
In each case, results are obtained for both the Hubbard and the quadratic dispersions.
A 4th-order polynomial function in $1/L$ fits well both dispersions, and the extrapolated results in continuum limit agree well with each other. The insets indicate that the coefficients on $1/L$ are negligible in both cases.
  }
\end{figure*}

The extrapolation to the continuum limit, 
for a fixed number of particles, 
must be consistent and independent of 
the type of kinetic energy dispersion.
For a two-body problem on the lattice, exact results can be obtained for large system sizes by 
mapping to a one-body problem in the center of mass system.
The results are shown in Fig.~\ref{fig:fss}(a), which 
fit well a 4th-order polynomial function in $1/L$. We see from the inset that the coefficient on the linear term  is zero within numerical precision.

We also show the finite size effect in the four-body problem from QMC, in Fig.~\ref{fig:fss}(b), reaching large lattice sizes. The same general behavior is seen as in the two-body problem.  
We have also studied the finite-size behavior of the BCS solution, finding similar trends but with different slopes.
In the many-body system, our QMC data are consistent with these observations as well. They are thus fitted 
 with a 4th-order polynomial function with a vanishing $1/L$ coefficient, as described in the main text.

\section{Equation of state data \label{appendix:eos}}
We list the data for the equation of state in Fig.~\ref{fig:energy}.  The QMC energy data are calculated by our branching random walk approach with BCS trial wave functions.

\begin{table}
\caption{Data of the equation of state in Fig.~\ref{fig:energy}. 
The interaction strength, given in the first column, are $\ln(a k_F)=y+\ln(2)/2$, with $y$ from 
$-0.75$ to $6$ (in increments of $0.25$ up to $y=2$, then increments of 0.5 up to $y=5$).
}. 
\begin{ruledtabular}
\begin{tabular}{cccc}
$\ln(a k_F)$ & $E_{QMC}/E_{FG}$ & Error bar & $E_{BCS}/E_{FG}$\\
\hline
   -0.403426     &    -5.512634      &      0.000619    &    -4.651252  \\
   -0.153426     &    -3.262997      &      0.000487    &    -2.427641  \\
    0.096574     &    -1.884889      &      0.000325    &    -1.078969  \\
    0.346574     &    -1.027841      &      0.000453    &    -0.260958  \\
    0.596574     &    -0.487667      &      0.000335    &     0.235190  \\
    0.846574     &    -0.137058      &      0.000272    &     0.536119  \\
    1.096574     &     0.096228      &      0.000203    &     0.718642  \\
    1.346574     &     0.256943      &      0.000167    &     0.829348  \\
    1.596574     &     0.371799      &      0.000162    &     0.896491  \\
    1.846574     &     0.456471      &      0.000141    &     0.937204  \\
    2.096574     &     0.521804      &      0.000173    &     0.961859  \\
    2.346574     &     0.572904      &      0.000111    &     0.976740  \\
    2.846574     &     0.647340      &      0.000103    &     0.990927  \\
    3.346574     &     0.700067      &      0.000067    &     0.997096  \\
    3.846574     &     0.737144      &      0.000128    &     0.997307  \\
    4.346574     &     0.767283      &      0.000099    &     0.997765  \\
    4.846574     &     0.793547      &      0.000068    &     1.000206  \\
    5.346574     &     0.813073      &      0.000053    &     1.000436  \\
    6.346574     &     0.842689      &      0.000036    &     1.000654 
\end{tabular}
\end{ruledtabular}
\end{table}

\section{Full digits for Table \ref{tab:fitparam} \label{appendix:fitparam}}
\begin{table}
\caption{\label{tab:fitparamfull} 
Final parameter values (full digits) in the parametrization [Eqs.~(\ref{eqn:fitcross}-\ref{eqn:fitbec})] of the exact EOS from QMC. 
}
\begin{ruledtabular}
\begin{tabular}{cccc}
&  $a^l_0$ &  -11.804127317953723 & \\
&  $a^l_1$ &  14.675499370762239  &\\
&  $a^l_2$ &  -4.855080880566919  &\\
&  $a_0$   &  -0.819842357425408  &\\
&  $a_1$   &  0.1273251139440354  &\\
&  $a_2$   &  0.0685123559420463  &\\
&  $a_3$   &  -0.014505432043856327&\\
&  $a_4$   &  -0.009191602440101383 &\\
&  $a_5$   &  0.004190575139056055   &\\
&  $a_6$   &  -0.0006367374265820822   &\\
&  $a_7$   &  3.431232818866204$\times 10^{-5}$  &\\
&  $a^r_2$ & -0.060852400057644876  &\\
&  $a^r_3$ &  0.36401186693517423   &\\
&  $a^r_4$ & -0.61531422724189
\end{tabular}
\end{ruledtabular}
\end{table}

\begin{acknowledgments}
We thank J.~Carlson 
for useful discussions. 
This research was supported by DOE (grant no.~DE-SC0008627), NSF 
(grant no.~DMR-1409510), and the Simons Foundation. Computing was carried out at the 
Oak Ridge Leadership Computing Facility at the Oak Ridge National Laboratory,
 which is supported by the Office of Science of the U.S. Department of Energy under Contract No. DE-AC05-00OR22725,
and at the computational facilities at the College of William and Mary.
\end{acknowledgments}

\bibliography{FG2D}

\begin{thebibliography}{47}
\bibitem[{FS-()}]{FS-corr-note}
\bibinfo{note}{We have applied finite-size corrections to our QMC results using
  the difference between the BCS energy for the same $N$ and at the TL. The
  correction is in general small (largest at large $x$) and vanishes with
  increasing $N$.}
\bibitem[{Par()}]{Param-EOS-BCS-a2--note}
\bibinfo{note}{Note that our $a^r_2$ has different sign from the DMC result
  \cite{PhysRevLett.106.110403}, but is consistent with analytic results of
  $-0.05908$ \cite{PhysRevB.45.10135,PhysRevA.90.053633}.}
\end{thebibliography}

\end{document}